\documentclass[aps,prx,twocolumn,floatfix,superscriptaddress,graphics,longbibliography]{revtex4-1}

\usepackage{amsmath,amssymb,graphicx,color,braket,makeidx,xcolor,subfigure, dsfont}
\usepackage{bm}
\usepackage[retainorgcmds]{IEEEtrantools}
\usepackage{graphicx}
\usepackage{mathrsfs}
\usepackage{amsmath}
\usepackage{amssymb}
\usepackage{color}
\usepackage{amsfonts}
\usepackage{amsthm}
\usepackage{times,txfonts}
\usepackage{nicefrac}
\usepackage[colorlinks=true,linkcolor=blue,urlcolor=blue,citecolor=blue,pdfusetitle]{hyperref}
\usepackage{framed}
\usepackage{yfonts}

\newtheorem*{definition*}{Def}
\newtheorem*{theorem*}{Theorem}
\newtheorem*{corollary*}{Corollary}
\newtheorem*{lemma*}{Lemma}
\newtheorem*{teorema*}{Teorema}

\renewcommand{\qedsymbol}{$\blacksquare$}
\providecommand{\bgreek}[1]{\mbox{\boldmath$#1$}}
\newcommand{\Ham}{\mathcal{H}}

\begin{document}

\title{Hyperaccurate thermoelectric currents}
\date{\today}
\author{Andr\'e M. Timpanaro}
\affiliation{Universidade Federal do ABC,  09210-580 Santo Andr\'e, Brazil}
\author{Giacomo Guarnieri}
\affiliation{Dahlem Center for Complex Quantum Systems,
Freie Universit\"{a}t Berlin, 14195 Berlin, Germany}
\author{Gabriel T. Landi}
\affiliation{Instituto de F\'isica da Universidade de S\~ao Paulo,  05314-970 S\~ao Paulo, Brazil.}

\begin{abstract}
Thermodynamic currents can fluctuate significantly at the nanoscale. 
But some currents fluctuate less than others.
Hyperaccurate currents are those which fluctuate the least, in the sense that they maximize the signal-to-noise ratio (precision). 
In this letter we analytically determine the hyperaccurate current in the case of a quantum thermoelectric, modeled by the Landauer-B\"uttiker formalism. 

\end{abstract}

\maketitle

{\bf \emph{Introduction--}} 
Fluctuations of the power grid can fry our electronics, a fact which is widely taken into account in the manufacturing of chips.
And the stride towards miniaturization, which is still a major technological drive, is constantly bringing these devices closer the nanoscale, where fluctuations are even more relevant.
Consequently, strategies for curbing them have become a focal point of an ever-growing research field, aimed at designing next-generation quantum devices. 
Whenever one considers a generic thermodynamic current (e.g. heat, work, charge, etc.) passing through such a small system, the relevant figure of merit is the signal-to-noise ratio (SNR), defined as 
\begin{equation}\label{SNR}
    \mathcal{S}_J = J^2/\Delta_J,
\end{equation}
with $J$ denoting the average current and $\Delta_J^2$ its corresponding variance. 

Some currents fluctuate more than others, however
(e.g. heat could perhaps fluctuate more than energy).
Recently, Busiello and Pigolotti introduced the concept of a \textit{hyperaccurate current}~\cite{Busiello2019} as the one which possesses the maximum SNR $\mathcal{S}_{\rm hyp}$ among all currents;  i.e., $\mathcal{S}_J \leqslant \mathcal{S}_{\rm hyp}$.
Their results hold for classical non-equilibrium steady-states of overdamped Langevin systems. 
In this case it has also been shown in~\cite{Barato2015,Gingrich2016,Pietzonka2015,Horowitz2019} that any SNR is always upper bounded by the entropy production rate $\sigma$, according to 
\begin{equation}\label{TUR}
    \mathcal{S}^{(\rm cl)}_J \leqslant \sigma/2, 
\end{equation}
where the superscript `$(\rm cl)$` is placed to emphasize that this holds for classical stochastic processes. 
This result, now famously known as thermodynamic uncertainty relation (TUR), establishes a tradeoff: it implies that to improve precision, one must pay the price in entropy production, which is a measure of dissipation~\cite{Polettini,Gingrich_2017,Pietzonka2017, VuPRR2020,PaneruPRE2020,Niggemann_2020}. 
Taken together, the hyperaccurate current and Eq.~\eqref{TUR} imply the sequence of bounds $\mathcal{S}^{(\rm cl)}_J \leqslant \mathcal{S}^{(\rm cl)}_{\rm hyp} \leqslant \sigma/2$. 
Moreover, it was proven in Ref.~\cite{Vu2019PRE} that, in the case of overdamped classical Langevin systems, the only current able to saturate Eq.~\eqref{TUR}, in principle, is the entropy production rate itself. 
In all other cases, the hyperaccurate current  provides a tighter estimate instead.

These results may fail at the nanoscale, however, where quantum effects become preponderant. Crucially, the TUR~\eqref{TUR} can  be violated in this case~\cite{Agarwalla2018,Ptaszynski2018a,Saryal2019,Guarnieri2019,Liu2019,Cangemi2020,
Ehrlich2021,Kalaee2021,Saryal2020,Liu2020,timpanaro2021precise}, i.e. $\mathcal{S}_J \nleqslant \sigma/2$.
This makes the determination of the hyperaccurate current much more important, as it would serve as the only available bound. 

In this Letter we address this issue in the context of quantum thermoelectrics described by the Landauer-B\"uttiker formalism.
For any given setup, we show how to analytically compute the hyperaccurate current $J_{\rm hyp}$, i.e. the current having the largest possible SNR $\mathcal{S}_{\rm hyp}$. 
Since TURs can be violated in thermoelectrics, this becomes particularly useful, as it provides the ultimate precision for any other current \begin{equation}\label{hyper_bound}
    \mathcal{S}_J \leqslant \mathcal{S}_{\rm hyp},
\end{equation}
hence in turn paving the way for the design of optimized next generation quantum devices.



\begin{figure}
    \centering
    \includegraphics[width=0.35\textwidth]{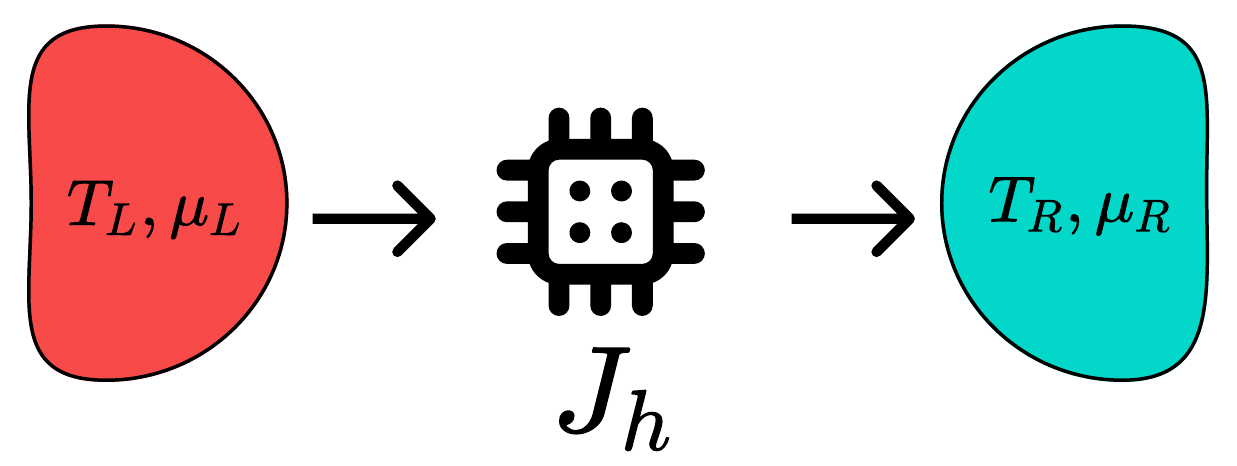}
    \caption{A quantum thermoelectric coupled to two baths at different temperatures and chemical potentials.}
    \label{fig:drawing}
\end{figure}

{\bf \emph{The problem--}}
We consider a quantum thermoelectric system, characterized by an energy-dependent transmission function $\mathcal{T}(\epsilon) \in [0,1]$. 
The system is coupled to two baths, kept at different temperatures $T_i$ and chemical potentials $\mu_i$, $i = L, R$ (Fig.~\ref{fig:drawing}). 
A generic current, within the Landauer-B\"uttiker formalism, has the form~\cite{Datta1997a}   
\begin{equation}\label{J}
    J_h = \int d\epsilon~ h(\epsilon)\mathcal{T}(\epsilon) \Delta f(\epsilon),
\end{equation}
where $\Delta f = f_L - f_R$ and $f_i(\epsilon) = (e^{\beta_i(\epsilon-\mu_i)}+1)^{-1}$ is the Fermi-Dirac occupation of bath $i$, and $h(\epsilon)$ defines the type of current in question. 
The particle current $J_N$ corresponds to $h(\epsilon) = 1$;  the energy current $J_E$ has $h(\epsilon) = \epsilon$; 
and the heat currents $J_{Q}^{L(R)}$ have $h(\epsilon) = J_E - \mu_i J_N$.
In principle, more complicated (non-linear) functions $h(\epsilon)$ are also allowed.
These are not typical in the thermoelectric scenario, and could be very difficult to measure. But, as we show in the Supplemental Material~\cite{SupMat}, they are nonetheless physically valid currents. 
We will therefore analyze both generic non-linear $h(\epsilon)$, and linear $h(\epsilon)$ (which lead to standard heat currents). 

Within the Levitov-Lesovik full counting statistics formalism~\cite{Levitov1993,Esposito2009}, the variance associated to any such current reads
\begin{equation}\label{var}
    \Delta_h = \int d\epsilon~ h(\epsilon)^2 \mathcal{T}(\epsilon)\left[f_L + f_R - 2f_R f_L - \mathcal{T}(\epsilon)\Delta f^2 \right].
\end{equation}
In~\cite{SupMat} it is shown that this holds for any $h(\epsilon)$.
The SNR~\eqref{SNR} is then $\mathcal{S}_h = J_h^2/\Delta_h$.

{\bf \emph{Main results--}}
Our first main result is the following: 

{\bf Theorem 1} (hyperaccurate): \emph{For a given transmission function $\mathcal{T}(\epsilon)$, and fixed $T_L, T_R, \mu_L, \mu_R$, the current $h(\epsilon)$ which maximizes the SNR is given by 
\begin{equation}\label{h_opt}
    h_{\rm hyp}(\epsilon) = \frac{\Delta f(\epsilon)}{g(\epsilon) + \Delta f(\epsilon)^2 \big[1 - \mathcal{T}(\epsilon)\big]},
\end{equation}
where $g = f_L(1-f_L) + f_R(1-f_R)$. 
The corresponding optimal SNR is given by 
\begin{equation}\label{snr_opt}
    \mathcal{S}_{\rm hyp} =\int  d\epsilon \frac{\Delta f(\epsilon)^2 \mathcal{T}(\epsilon)}{g(\epsilon) + \Delta f(\epsilon)^2 \big(1 - \mathcal{T}(\epsilon)\big)} .
\end{equation}
The SNR is insensitive to a change $h \to \alpha h$, so
Eq.~\eqref{h_opt} is only defined up to a scale.
}

\emph{Proof:} Since the problem is invariant under a change of scale, to maximize $J_h^2/\Delta_h$, we can without loss of generality fix the scale so that $J_h \equiv 1$.
The problem then becomes that of minimizing $\Delta_h$ over $h(\epsilon)$, subject to the constraint that $J_h = 1$. 
This is a convex optimization problem and can thus be treated with standard Lagrange multipliers. We introduce the functional $G[h(\epsilon)] = \Delta_h + \lambda (J_h - 1)$, where $\lambda$ is a Lagrange multiplier. 
Extremizing over $h(\epsilon)$, using standard variational calculus,  yields Eq.~\eqref{h_opt}. 
And substituting $h_{\rm hyp}$ in Eqs.~\eqref{J} and~\eqref{var} yields~\eqref{snr_opt}. \qedsymbol

Next, we consider the case where $h(\epsilon)$ is linear, of the form $h(\epsilon)  = a \epsilon + b$, for constants $a$, $b$: 

{\bf Theorem 2} (linear hyperaccurate): \emph{
Restricting to linear functions $h(\epsilon)$, the hyperaccurate current has the form 
\begin{equation}
    h_{\rm lhyp}(\epsilon)  = (J_E \Delta_N^2 - J_N \mathcal{C}) \epsilon + (J_N \Delta_E^2 - J_E \mathcal{C}), 
\end{equation}
where $J_{E(N)}$ are the energy and particle currents, $\Delta_{E(N)}$ the corresponding variances and $\mathcal{C} = \Delta_{h = \sqrt{\epsilon}}$ is the covariance between energy and particle currents. The corresponding SNR  reads
\begin{equation}\label{linear_snr_opt}
    \mathcal{S}_{\rm lhyp} = \frac{\mathcal{S}_E + \mathcal{S}_N - 2 f \sqrt{\mathcal{S}_E \mathcal{S}_N}}{1-f^2},
\end{equation}
where $f = \mathcal{C}/\sqrt{\Delta_E \Delta_N}$ is the Fisher correlation of the energy and particle currents.
For any other linear current $h = a \epsilon + b$ (including particle and energy) it thus follows that 
\begin{equation}\label{bound_orders}
\mathcal{S}_{h} \leqslant \mathcal{S}_{\rm lhyp} \leqslant
    \mathcal{S}_{\rm hyp}.
\end{equation}
}

\emph{Proof:} The proof follows directly by  plugging $h(\epsilon) = a \epsilon + b$ in Eq.~\eqref{SNR}. The result is a function $S(a,b)$ which can be maximized by standard methods. 
In doing so, it is convenient to notice that, e.g., $J_{a\epsilon + b} = a J_E + b J_N$, where $J_E = J_{h = \epsilon}$ and $J_N = J_{h = 1}$. 
Similar formulas hold for the variance in Eq.~\eqref{var}. The only subtle point is that, when expanding $h(\epsilon)^2 = (a\epsilon+b)^2$ in the variance, there will be a term of the form $2ab \epsilon$ which can be equivalently interpreted either as $2 a b \Delta_{h = \sqrt{\epsilon}}$. 
Alternatively, one can also interpret $ \Delta_{h = \sqrt{\epsilon}} \equiv \mathcal{C}$ , as the covariance between the energy and particle currents~\footnote{
This can be proven, for instance, starting from the joint cumulant generating function of energy and particle currents; c.f.~\cite{Liu2019}. 
}. \qedsymbol

We mention in passing that the linear hyperaccurate current is not at all related to the linear response regime (infinitesimal temperature and chemical potential gradients), and can very well be defined arbitrarily far from linear response.




{\bf \emph{Linear response--}} In order to gain further insight into the physical meaning of Theorem 1,
we study small gradients by parametrizing
\begin{IEEEeqnarray*}{rCCCrCl}
\beta_L &=& \beta + \delta_\beta/2, &\qquad& \beta_R &=& \beta - \delta_\beta/2, 
\\[0.2cm]
\beta_L \mu_L &=& \beta \mu + \delta_{\beta\mu}/2,
&\qquad&
\beta_R \mu_R &=& \beta \mu - \delta_{\beta\mu}/2.
\end{IEEEeqnarray*}
A series expansion in powers of 
$\delta_\beta$ and $\delta_{\beta\mu}$ then yields 
$g(\epsilon) \simeq 2 f(1-f)$ 
and 
$\Delta f \simeq (\delta_{\beta \mu}- \delta_\beta \epsilon)f(1-f)$, where
$f = (e^{\beta(\epsilon-\mu)}+1)^{-1}$ is the Fermi-Dirac distribution associated to the mean temperature and chemical potential.
As a consequence, the optimal current~\eqref{h_opt} reduces to 
\begin{equation}\label{h_opt_linear_response}
    h_{\rm hyp} = \delta_\beta \epsilon - \delta_{\beta \mu},
\end{equation}
once again defined up to a scale factor.
In this case, therefore, the hyperaccurate current is itself a linear current,  so the results from Theorems 1 and 2 coincide [Eqs.~\eqref{h_opt} and~\eqref{h_opt_linear_response}]. 

Inserting~\eqref{h_opt_linear_response} in~\eqref{J} yields a current $J_{\rm hyp} \propto J_E - \epsilon^* J_N$, where $\epsilon^* = \delta_{\beta\mu}/{\delta_\beta}$. 
The corresponding SNR~\eqref{snr_opt} becomes 
\begin{equation}
    \mathcal{S}_{\rm hyp} = \frac{1}{2} \int d\epsilon~ \mathcal{T}(\epsilon)(\delta_{\beta \mu} - \delta_\beta \epsilon)^2 f(1-f).
\end{equation}
One may now verify that this is exactly $\sigma/2$, 
where $\sigma = - \delta_\beta J_E + \delta_{\beta\mu} J_N$ is the entropy production rate~\cite{Yamamoto2015a}. 
When, \emph{in the regime of linear response, the hyperaccurate current is proportional to the entropy production rate and hence saturates the TUR~\eqref{TUR}}. 
Away from linear response, this breaks down completely.

\begin{figure}
    \centering
    \includegraphics[width=0.4\textwidth]{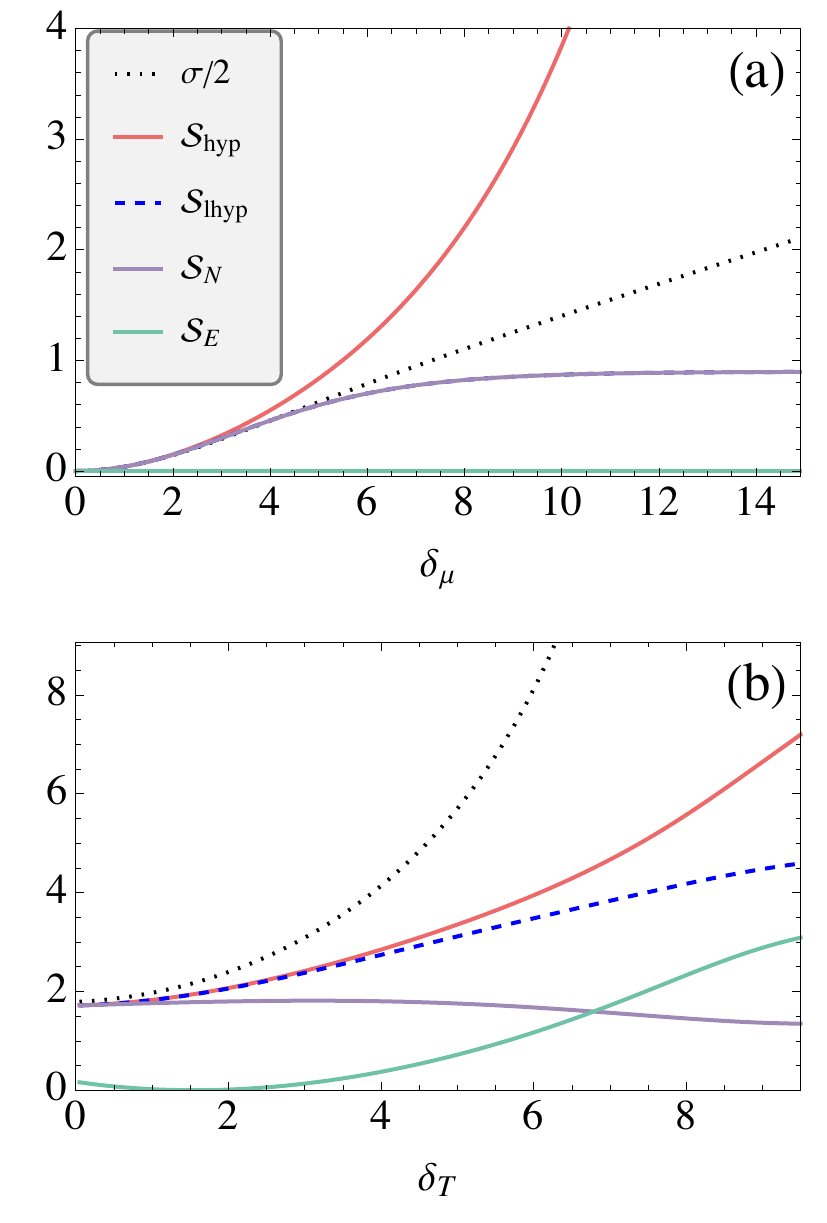}
    \caption{Signal-to-noise ratio (SNR) for the double quantum dot transmission function~\eqref{bridge}. 
    The green and purple curves correspond to the SNR for energy and particles, $\mathcal{S}_{E(N)}$,  respectively. 
    The hyperaccurate bound~\eqref{snr_opt} is shown in red, the linear SNR~\eqref{linear_snr_opt} in blue-dashed, and $\sigma/2$ [Classical TUR~\eqref{TUR}] in gray-dotted.
    (a) As a function of the chemical potential gradient $\mu_R = -\mu_L = \delta_\mu/2$, with $\Gamma = \Omega = 0.1$, $T_L = 0.8$, $T_R = 1$.
    (b) As a function of the temperature gradient $T_L = T_0 - \delta_T/2$ and $T_R = T_0 + \delta_T/2$, with $T_0 = 5$, $\Gamma = 6.5$, $\Omega = 10$ and $\mu_L = 6$, $\mu_R = 0$.
    Notice that these plots refer only to the SNR, not the actual average currents. 
    In (a) the purple curve ($\mathcal{S}_N$) practically overlaps with $S_{\rm lhyp}$ (blue-dashed). 
    }
    \label{fig:example}
\end{figure}

{\bf \emph{Example--}} As an example, we consider a resonant double quantum dot with transmission function given by
\begin{equation}\label{bridge}
    \mathcal{T}_d(\epsilon) = \frac{\Gamma^2 \Omega^2}{|(\epsilon-+i\Gamma/2)^2-\Omega^2 |^2},
\end{equation}
where $\Gamma$ denotes the bath coupling strength and $\Omega$ the inter-dot hopping constant. 
Fig.~\ref{fig:example} shows the behavior of the SNRs for the energy and particle currents, $\mathcal{S}_E$ (green) and $\mathcal{S}_N$ (purple). These are also compared with the hyperaccurate bound~\eqref{snr_opt} (red) and the linear hyperaccurate bound~\eqref{linear_snr_opt} (blue-dashed). 
In (a) we set $\Gamma = \Omega = 0.1$, $T_L = 0.8$, $T_R = 1$, and vary the chemical potential gradient by setting $\mu_R = -\mu_L = \delta_\mu/2$. 
In Fig.~\ref{fig:example}(b) we set $\Gamma = 6.5$, $\Omega = 10$, $\mu_L = 6$, $\mu_R = 0$ and vary the temperature gradient by setting $T_L = T_0 - \delta_T/2$ and $T_R = T_0 + \delta_T/2$, with $T_0 = 5$. 
In both cases all the SNRs are bounded according to Eq.~\eqref{bound_orders}. For completeness, we also plot the classical TUR bound $\sigma/2$ [Eq.~\eqref{TUR}] in gray-dotted. 
But since this bound can be violated in quantum thermoelectrics, there is no fixed order relation between it and the other SNRs (e.g. $\sigma/2 < \mathcal{S}_{\rm hyp}$ in Fig.~\ref{fig:example}(a) and $\sigma/2 > \mathcal{S}_{\rm hyp}$ in (b)).

It is finally worth emphasizing that Fig.~\ref{fig:example} refers to the SNR, and not to the respective currents. 
In Fig.~\ref{fig:example}(a) the energy current fluctuates very little and the particle current SNR matches very closely the linear bound~\eqref{linear_snr_opt}. 
Conversely, in Fig.~\ref{fig:example}(b) the SNR of $\mathcal{S}_E$ and $\mathcal{S}_N$ are similar in magnitude, and both are strictly smaller than $\mathcal{S}_{\rm lhyp}$.
Moreover, in this case it turns out that $\mathcal{S}_{\rm lhyp}$ is not far from the non-linear bound $\mathcal{S}_{\rm hyp}$ [Eq.~\eqref{snr_opt}], whereas in Fig.~\ref{fig:example}(a) their mismatch can be significant. 

{\bf \emph{Significance--}} 
The hyperaccurate current~\eqref{h_opt} is generally a non-linear function of energy. 
This therefore may not entail it a very practical physical interpretation. 
Instead, its significance lies in the fact that it provides a universal bound, which must be satisfied by the SNR of any other current. 
That is, for any $\mathcal{S}_h$, it follows that 
$\mathcal{S}_h \leqslant \mathcal{S}_{\rm hyp}$ [Eq.~\eqref{hyper_bound}].
If the current in question is linear ($h = a \epsilon + b$), one also has an even tighter bound bound
$\mathcal{S}_h \leqslant \mathcal{S}_{\rm lhyp}$ [Eq.~\eqref{bound_orders}]. 
These bounds can thus be used as guidelines in devising strategies for curbing the fluctuations of different currents. 
For  linear currents, in particular, $\mathcal{S}_{\rm lhyp}$ can be directly associated to experimentally measurable quantities. 
In classical systems this kind of upper bound would be given instead by the TUR~\eqref{TUR}; that is,  the entropy production would determine the ultimate precision. 
But for thermoelectrics the TUR can be violated (e.g. Fig.~\ref{fig:example}), so one must use Eq.~\eqref{hyper_bound} instead.  
We believe this represents the main significance of the hyperaccurate currents within a quantum thermoelectric scenario.

\emph{\textbf{Acknowledgements--}} 
The authors acknowledge fruitful discussions R. Andrade, A. M. Cunha, A. Teixeira, I. Ferreira, I. Queiroz, M. Grael, K. Kunze, L. Bus\'{a}, M. Jacobs and G. Tamberi.
GTL acknowledges the financial support of the  S\~ao Paulo Funding Agency FAPESP (Grants No. 2017/50304-7, 2017/07973-5 and 2018/12813-0), the Gridanian Research Council (GRC), and the Brazilian funding agency CNPq
(Grant No. INCT-IQ 246569/2014-0). G. G. acknowledges support from FQXi and DFG FOR2724 and also from the European Union Horizon 2020 research and innovation programme under the Marie Sklodowska-Curie grant agreement No. 101026667.

\bibliography{library,libsup,newlib}

\pagebreak
\widetext

\newpage 
\begin{center}
\vskip0.5cm
{\Large Supplemental Material}
\end{center}
\vskip0.4cm

\setcounter{section}{0}
\setcounter{equation}{0}
\setcounter{figure}{0}
\setcounter{table}{0}
\setcounter{page}{1}
\renewcommand{\theequation}{S\arabic{equation}}
\renewcommand{\thefigure}{S\arabic{figure}}

In this supplemental material, we prove that the variance associated to an arbitrary current $h(\epsilon)$ is indeed given by Eq.~\eqref{var} of the main text. 

\section{Basic notation and the two-point measurement scheme}

We consider a quantum system with initial state $\hat{\rho}$, evolving unitarily with evolution operator $\hat{U}(t,t_0)$. 
Let
$\hat{O}(t) = \sum_{o_t} o_t \hat{\Pi}_{o_t} $ denote a generic observable, with eigenvalues $o_t$ and projectors $\hat{\Pi}_{o_t}$.
Throughout, we will repeatedly make use of the spectral theorem, which states that  given an operator $\hat{O}$, as above, a generic function of it will have the following decomposition 
\begin{equation}\label{ST}
  f(\hat{O}_t) = \sum_{o_t} f(o_t) \hat{\Pi}_{o_t}.
\end{equation}

Within the two-point measurement (TPM) scheme, the probability of observing $o_{t_i}$ at time $t_i$ and then $o_t$ at time $t$ is given by 
\begin{equation}\label{probdistrFCS}
    p_t(\Delta o) = \sum_{o_i,o_t} \delta\left(\Delta o - (o_t-o_i)\right) \mathrm{Tr}\left[\hat{\Pi}_{o_t}\hat{U}(t,t_i)\hat{\Pi}_{o_i}\hat{\rho}(t_i)\hat{\Pi}_{o_i}\hat{U}^{\dagger}(t,t_i)\hat{\Pi}_{o_t}\right],
\end{equation}
with $\delta(.)$ denoting the Dirac's delta function and $\hat{\rho}(t_i)$ being the initial state.
We also define the cumulant generating function (CGF)
\begin{equation}\label{eq:CGF1}
    \mathcal{G}(\eta)_t = \ln \int d(\Delta o) e^{-i\eta \Delta o} p_t(\Delta o).
\end{equation}
From this, the variance can be obtained by differentiating over $\eta$: $\langle(\Delta o_t)^2\rangle = (-i)^2\partial_{\eta^2}\mathcal{G}(\eta)_t|_{\eta=0}$.
Assuming that the initial density matrix $\hat{\rho}(t_i)$ commutes with  $\hat{O}_{t_i}$, it is also possible to 
to rewrite the CGF as
\begin{equation}\label{CGF2}
    \mathcal{G}_t(\eta) = \ln \mathrm{Tr}\left[e^{i\eta \hat{O}_t}\hat{U}(t,t_i) e^{i\eta \hat{O}_i}\hat{\rho}(t_i)\hat{U}^{\dagger}(t,t_i)\right].
\end{equation}

\section{Full counting statistics for hyperaccurate fermionic current in quantum thermoelectric systems}

We now analyze the CGF in the case of non-interacting fermions.
We consider a composite system, with total Hamiltonian 
\begin{equation}
    \hat{\Ham} = \hat{\Ham}_S + \sum_{\alpha} (\hat{\Ham}_{\alpha} + \hat{V}_{S\alpha}), \,\, (\alpha = L,R).
\end{equation}
The first term is the Hamiltonian of the system, which is taken to be described by a quadratic Hamiltonian in a set of fermionic operators $c_s$; that is,  $\hat{\Ham}_{S} = \sum_{s} \epsilon_s \hat{c}^{\dagger}_s\hat{c}_s$. 
This system functions as a junction, connecting a left and right bath, with Hamiltonians 
$\hat{\Ham}_{L(R)} = \sum_{l(r)} \epsilon_{l(r)} \hat{c}^{\dagger}_{l(r)}\hat{c}_{l(r)}$. 
Finally $\hat{V}_{SL(R)} = \sum_{s,l(r)} \left( J_{sl(r)} \hat{c}^{\dagger}_s\hat{c}_{l(r)} + h.c.\right)$ is the corresponding interaction. 
The above creation and annihilation operators for the system and the leads are fermionic operators each satisfy their own standard anti-commutation relations $\lbrace \hat{c}_j,\hat{c}^{\dagger}_k \rbrace = \delta_{jk}$, $\hat{c}_j\hat{c}_k = \hat{c}^{\dagger}_j\hat{c}^{\dagger}_k = 0$.

Usually, the observable that is considered is either the particle number $\hat{N}_L$ or the energy $\hat{\Ham}_L$ of one of the reservoirs, say the left one for concreteness. 

But here we will be interested in measuring a generic observable of the form $\hat{O}_{h} = h(\hat{\Ham}_L)$.  By virtue of the spectral theorem Eq.~\eqref{ST}, its spectral decomposition in the Fock basis is  given by
\begin{equation}\label{caveat}
    \hat{O}_{h} = h(\hat{\Ham}_L) = \sum_l h(\epsilon_l) \hat{c}^{\dagger}_l\hat{c}_l.
\end{equation}
Thus, by construction, this observable does not have coherences between different Fock states, and commutes with both the Hamiltonian and the particle number operator of the left reservoir.
In order to be able to apply the TPM formalism of the previous Section, one finally needs to consider an initial state which commutes with the desired observable,  $\hat{O}_{h}$. 
We take, as is standard, the assumption that the interaction $\hat{V}_{SL}+\hat{V}_{SR}$ are adiabatically switched on from $t_i = - \infty$, so that the  initial state has the form 
$\hat{\rho}(-\infty) = \hat{\rho}_S(-\infty)\bigotimes_{\alpha=L,R} Z_{\alpha}^{-1} \exp\left[-\beta_\alpha(\hat{\Ham}_{\alpha}-\mu_{\alpha}\hat{N}_{\alpha})\right]$. We remark that the choice of the initial state for the central junction system is immaterial in the steady-state after the thermodynamic limit is performed in the baths~\cite{Fujii2013,tasaki2003fundamental}. We will therefore choose $\hat{\rho}_S(-\infty) = Z_S^{-1} e^{-\beta_S \left(\hat{\Ham}_S-\mu_S\hat{N}_S\right)}
$ from now on.

Taking into account that $t_i = -\infty$, we are therefore led to the following expression for the CGF of the hyperaccurate observable
\begin{align}\label{eq:CGF3}
    &\mathcal{G}_t(\eta) = \ln \mathrm{Tr}\left[\hat{U}^{\dagger}(t,-\infty) e^{i\eta \hat{O}_{h}}\hat{U}(t,-\infty) e^{i\eta \hat{O}_{h}}\hat{\rho}(-\infty)\right]\notag\\
   &= \ln \mathrm{Tr}\left[\hat{U}^{\dagger}(t,-\infty) e^{i\eta \sum_l h(\epsilon_l) \hat{c}^{\dagger}_l\hat{c}_l} \hat{U}(t,-\infty) e^{i\eta \sum_l h(\epsilon_l) \hat{c}^{\dagger}_l\hat{c}_l}\hat{\rho}(-\infty)\right]
\end{align}
Finally, a useful reformulation of the above result was found by Klich ~\cite{klich2003elementary}, which allowed to express the CGF as a determinant of second-quantization operators restricted to their single-particle subspace. The latter can be summarized as follows: let $\hat{X} = \sum_{ij} X_{ij} \hat{c}^{\dagger}_i\hat{c}_j$ be a fermionic bilinear operator and $\mathbf{X}$ denote the matrix with entries $X_{ij}$; then the following identity can be proven
\begin{equation}\label{Klich1}
    \mathrm{Tr}\left[e^{\hat{X}}\right] = \mathrm{det}\left[1+e^{\mathbf{X}}\right].
\end{equation}
The same holds true for any product of fermionic bilinear operators, i.e. $\mathrm{Tr}\left[e^{\hat{X}}e^{\hat{Y}}\right] = \mathrm{det}\left[1+e^{\mathbf{X}}e^{\mathbf{Y}}\right]$. 
For the problem at hand, one has that
\begin{equation}
    \mathbf{O}_{h} = \mathrm{diag}\left(h(\epsilon_l)\right)\oplus \mathbf{1}_{SR}
\end{equation}
and 
\begin{equation}
    \bgreek{\rho}(-\infty) = \bigoplus_{\alpha=L,R,S} \mathrm{diag}\left(n_\alpha(\epsilon_{\alpha})\right),
\end{equation}
where $n_\alpha(\epsilon_{\alpha}) = \left[1+e^{-\beta_{\alpha}(\epsilon_{\alpha}-\mu_{\alpha})}\right]^{-1}$.
The application of Eq.~\eqref{Klich1} to Eq.~\eqref{eq:CGF3} then leads to
\begin{equation}\label{eq:CGFgeneral}
    \mathcal{G}_t(\eta) 
    = \mathrm{det}\left[\mathbf{1}_{LSR} - \bgreek{\rho}(-\infty) + \bgreek{\rho}(-\infty)\left( \mathbf{U}^{\dagger}(t,-\infty) e^{i\eta \,\mathbf{O}_{h}}\mathbf{U}(t,-\infty) e^{i\eta \,\mathbf{O}_{h}} \right)\right].
\end{equation}
Eq.~\eqref{eq:CGFgeneral} represents the general expression for the current statistics of thermoelectric devices consisting of non-interacting fermions, at any given time, and without any approximation. In the following Section, we will consider the case of time-independent elastic scattering, also known as Landauer-B\"{u}ttiker regime, where the CGF reduces to the so-called Levitov-Lesovik expression.

\section{Elastic scattering}

Let us from now on assume that the scattering region is of similar size or smaller compared to the relaxation length-scales of the electrons (coherent scattering).
Furthermore, we will focus our attention to elastic scattering, which implies that each electron moves from one reservoir to the other as a plane wave with energy $\epsilon$. This condition is equivalent to requiring energy conservation and the validity of a large deviation principle, which is an assumption usually satisfied at the steady-state regime, whenever the size of the central system is negligible compared to the size of the baths (for which a thermodynamic limit is performed)~\cite{Esposito2009}. 
In this case, electrons with different energies thus contribute to particle-, energy- or any other such observable's (e.g. the hypercurrent's) statistics independently and thus $\mathcal{G}_t(\eta) = \int d\epsilon~ \mathcal{G}_t(\eta,\epsilon)$.
Furthermore, one can describe the above problem by taking the long-time limit $t\to+\infty$ in Eq.~\eqref{eq:CGF3}, and introduce the so-called \textit{scattering matrix} $\hat{S} = \lim_{t\to\infty} e^{i(\hat{\Ham}_L+\hat{\Ham}_R) t} \hat{U}(t,-t) e^{-i(\hat{\Ham}_L+\hat{\Ham}_R) t}$.
Exploiting the above considerations, the CGF Eq.~\eqref{eq:CGFgeneral} then reduces to
\begin{equation}\label{eq:LL}
    \chi(\eta) := \lim_{t\to\infty}\frac{1}{t}\mathcal{G}_t(\eta) = \int d\epsilon~ \ln\mathrm{det}\left[ \mathbf{1}_2 - \mathbf{n}_F(\epsilon) 
    + \mathbf{n}_F(\epsilon) \left(\mathbf{S}^{\dagger}(\epsilon) \mathbf{X}(\epsilon)\mathbf{S}(\epsilon)\right)\mathbf{X}^{\dagger}(\epsilon)\right],
\end{equation}
where $\mathbf{1}_2 = \begin{pmatrix}1&0\\0&1\end{pmatrix}$, and finally where we have introduced the two matrices
\begin{equation}
\mathbf{n}_F(\epsilon) = \begin{pmatrix}\left[1+e^{-\beta_L(\epsilon-\mu_L)}\right]^{-1}&0\\0&\left[1+e^{-\beta_R(\epsilon-\mu_R)}\right]^{-1}\end{pmatrix},
\qquad
\quad
\mathbf{X}(\epsilon) \equiv \begin{pmatrix}e^{-i\eta h(\epsilon)}&0\\0&1\end{pmatrix}.
\end{equation}
Upon expressing the scattering matrix in terms of reflection and transmission coefficients of the left and right leads $\mathbf{S}(\epsilon) = \begin{pmatrix}\mathrm{r}_{LL}(\epsilon) & \mathrm{t}_{LR}(\epsilon) \\ \mathrm{t}_{RL}(\epsilon)&\mathrm{r}_{RR}(\epsilon)\end{pmatrix} $, and making use of the following relations: \textit{(i)} $\mathrm{det}\left[\mathbf{S}\right]=1$, \textit{(ii)} $|\mathrm{r}_{LL}(\epsilon)|^2  + |\mathrm{t}_{LR}(\epsilon)|^2 = 1$, \textit{(iii)} $|\mathrm{r}_{RR}(\epsilon)|^2  + |\mathrm{t}_{RL}(\epsilon)|^2 = 1$ and \textit{(iv)}
$|\mathrm{t}_{LR}(\epsilon)|^2 = |\mathrm{t}_{RL}(\epsilon)|^2 \equiv \mathcal{T}(\epsilon) $, allows to finally arrive at the Levitov-Lesovik expression for the desired current
\begin{equation}
    \chi(\eta) = \int d\epsilon~ \ln\left\{1+\mathcal{T}(\epsilon)\left[(e^{h(\epsilon)\eta}-1)n_{L}(\epsilon)(1-n_{R}(\epsilon))+(e^{-h(\epsilon)\eta}-1)n_{R}(\epsilon)(1-n_{L}(\epsilon))\right)\right\}.
\end{equation}
From this, Eqs.~\eqref{J} and~\eqref{var} of the main text follow directly by taking the appropriate derivatives. 
This calculation therefore allows us to conclude that, indeed, it is in principle possible to measure generic currents, given by some function $h(\epsilon)$. 
The caveat to do so is to measure an operator of the form~\eqref{caveat}, interpreted in terms of the spectral theorem.

\bibliography{libsup}
\end{document}